\documentclass[%
 reprint,
 showpacs,
 amsmath,amssymb,
prb,
]{revtex4-1}
\usepackage{dcolumn}
\usepackage[dvipdfmx]{graphicx}
\usepackage{subfigure,color}
\usepackage{mathrsfs}
\usepackage{accents}

\makeatletter
\def\btt#1{\texttt{\@backslashchar#1}}
\DeclareRobustCommand\bblash{\btt{\@backslashchar}} \makeatother

\begin{document}

\title{Paramagnetic instability of small topological superconductors}

\author{Shu-Ichiro Suzuki$^1$ 
and Yasuhiro Asano$^{1,2}$}
\affiliation{$^1$Department of Applied Physics,
Hokkaido University, Sapporo 060-8628, Japan}
\affiliation{$^2$Center for Topological Science \& Technology,
Hokkaido University, Sapporo 060-8628, Japan}

\date{\today}

\begin{abstract}
The diamagnetism is an essential property of all superconductors. 
However, we will show that
small topological (or unconventional) superconductors can be intrinsically paramagnetic 
by solving the quasiclassical Eilenberger equation and 
the Maxwell equation self-consistently on two-dimensional superconducting disks 
in weak magnetic fields.
Because of the topologically nontrivial character of the wave function, 
the unconventional superconductors host the zero-energy surface Andreev bound states, 
which always accompany so-called odd-frequency Cooper pairs. 
The paramagnetic property of the odd-frequency pairs 
explains the paramagnetic response of the disks at low temperature.
\end{abstract}


\maketitle

\section{introduction}
The Meissner effect is a fundamental property of superconductors as shown in standard 
textbooks~\cite{tinkham}. The response of superconductors is usually diamagnetic because
a superconductor excludes weak enough magnetic fields from its interior. 
The anomalous paramagnetic response, however, has been observed 
in small disks of metallic superconductor~\cite{Thom1,Geim1}, 
small high-$T_c$ compounds~\cite{Brau1,Schl1,walter}, 
 and mesoscopic proximity structures~\cite{Visa1,Mota1}.
The spatial inhomogeneity of the magnetic property is a key feature to realize the 
paramagnetic phase.
In metallic superconductors, the inhomogeneous distribution of magnetic fields~\cite{koshelev} and
the formation of giant vortex are responsible for the paramagnetic Meissner effect
PME~\cite{moshcalkov}. 
The presence of the $\pi$ junctions is also pointed out as an origin of PME in a network 
of Josephson junction~\cite{dominguez}. 
In unconventional superconductors (USs), on the other hand, an experiment~\cite{walter} 
has shown the decrease of the pair density with decreasing temperature, which 
suggests a peculiar mechanism of the PME unique to the USs.
As a result of the topological nature in the wave function, the USs 
have the topologically protected surface Andreev bound states (ABSs) 
at the zero-energy~\cite{buchholtz,hara,hu,tanaka95,sato}.
So far theoretical studies have shown that 
the magnetic response at the (110) surface of high-$T_c$ superconductor is nonlinear~\cite{yip,Zare1}
and paramagnetic~\cite{fogelstrom,barash,Hig1,lofwander} due to the ABSs. 
The paramagnetic response has been mainly explained in terms of the energetics of the 
ABSs. Weak magnetic fields shift the energy of the surface ABSs away from the Fermi level
and decrease the total energy of superconductor, which leads to the paramagnetic response 
or the paramagnetic instability. 
However, there is an important open question: what carries the large paramagnetic supercurrent?
By addressing this issue, we will conclude that 
the magnetic properties of USs 
are intrinsically inhomogeneous and 
that small USs can be paramagnetic 
at low temperature.
 
The electric current in equilibrium has two contributions, (i.e., 
$\boldsymbol{j}=\boldsymbol{j}_{\textrm{pq}} + \boldsymbol{j}_{\textrm{A}}$).
The quasiparticle current $\boldsymbol{j}_{\textrm{pq}}$ due to the spatial phase gradient of 
the wave function is paramagnetic, whereas $\boldsymbol{j}_{\textrm{A}}=-ne^2\boldsymbol{A}/mc$ is
 diamagnetic. In a normal metal, $\boldsymbol{j}_{\textrm{pq}}$ cancels 
$\boldsymbol{j}_{\textrm{A}}$ because the phase of an electron is not rigid at all~\cite{AGD}.
In a superconductor, on the other hand, the phase rigidity of superconductivity drastically suppress 
the spatial gradient of phase, which leads to 
$\boldsymbol{j}_{\textrm{qp}}=0$. As a result, a superconductor shows the perfect diamagnetism. 
In contrast to excited quasiparticles \textsl{above} the superconducting gap, 
the quasiparticles \textsl{below} the gap have the phase rigidity 
because they are the shadow of Cooper pairs. 
In fact, a normal metal attaching to a metallic superconductor shows the diamagnetic 
Meissner effect~\cite{belzig}. 
This phenomenon is explained by two different but equivalent pictures: the penetration of 
a Cooper pair into the normal metal (proximity effect) and the Andreev reflection of 
a quasiparticle \textsl{below} the gap.  
The appearance of the surface ABSs is a direct result of 
the coherent Andreev reflections of a quasiparticle at the Fermi level~\cite{ya04}.
Therefore such phase-rigid quasiparticles at the ABS cannot carry the large paramagnetic current.
\par
\begin{figure}[b]
  \includegraphics[width=0.47\textwidth]{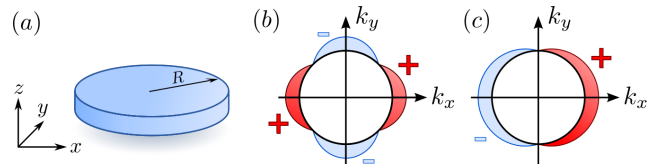}
  \caption{(a) The schematic figure of a superconducting disk. 
           The pair potentials in momentum space are illustrated for
           the $d$\, wave symmetry in (b) and for the $p$\, wave symmetry in (c). }
  \label{fig:schi}
\end{figure}

In this paper, we theoretically study the spatial distribution of magnetic fields and that of electric currents
on small two-dimensional superconducting disks with unconventional pairing symmetry such as spin-singlet $d$ wave 
and spin-triplet $p$ wave. There are several $d$ wave superconductors in organic compounds and heavy fermionic 
materials in addition to high-$T_c$ cuprates. Recently, the effective Hamiltonian 
for superconducting states in nanowires~\cite{Lutchyn2010,Oreg} has shown to be unitary equivalent to 
that for $p_x$ wave superconducting states~\cite{asano13}.
The simulation at least two-dimensional 
system is necessary to evaluate the magnetic susceptibility quantitatively because the 
$d$ and $p$ wave pair potentials are anisotropic in real space. 
We solve the Eilenberger equation for the quasiclassical 
Green function and the Maxwell equation for magnetic fields self-consistently. 
The self-consistency of pair potential and magnetic field is necessary to 
regularize the nonlinear property in the magnetic response~\cite{yip,Zare1}.
The solution of the Green function near the disk edge shows the presence of the odd-frequency Cooper pairs.
The odd-frequency pairs 
have paramagnetic property~\cite{Asa1,yoko1,melnikov12} because of their \textsl{negative pair density}. 
The calculated results of the magnetic susceptibility suggest the PME in small 
USs. We conclude that the odd-frequency Cooper pairs carry the large 
paramagnetic current and causes the paramagnetic response of small superconducting disks.

\section{formulation}
Let us consider a superconducting disk in two-dimension as shown in Fig.~\ref{fig:schi},  
where $R$ is the radius of the disk. 
We assume that the disk is in the clean limit and its surface is specular enough.
To analyze the superconducting states in equilibrium, 
we solve the Eilenberger equation~\cite{Eilen},
\begin{align}
& i\hbar v_F\hat{\boldsymbol{k}} \cdot 
\boldsymbol{\nabla}_{\boldsymbol{r}} \, \check{{g}}
+\left[ \check{H}, \check{g} \right]=0,\label{eilenberger_m2}\\
&\check{H}(\boldsymbol{r},{\boldsymbol{k}},i\omega_n)
=\left[\begin{array}{cc}
\hat{\xi}(\boldsymbol{r},{\boldsymbol{k}},i\omega_n) & 
\hat{\Delta}(\boldsymbol{r},{\boldsymbol{k}}) \\
\undertilde{\hat{\Delta}}(\boldsymbol{r},{\boldsymbol{k}}) & 
\undertilde{\hat{\xi}}(\boldsymbol{r},{\boldsymbol{k}},i\omega_n)
\end{array}\right], \\
&\check{g}(\boldsymbol{r},{\boldsymbol{k}},i\omega_n)
=\left[\begin{array}{cc}
\hat{g}(\boldsymbol{r},{\boldsymbol{k}},i\omega_n) & \hat{f}(\boldsymbol{r},{\boldsymbol{k}},i\omega_n) \\
-\undertilde{\hat{f}}(\boldsymbol{r},{\boldsymbol{k}},i\omega_n) & 
-\undertilde{\hat{g}}(\boldsymbol{r},{\boldsymbol{k}},i\omega_n)
\end{array}\right],\label{g_def}\\
&\hat{\xi}(\boldsymbol{r},{\boldsymbol{k}},i\omega_n)=i\omega_n 
+ (ev_F/ c){\boldsymbol{k}}\cdot \boldsymbol{A}(\boldsymbol{r}),
\end{align}
where ${\boldsymbol{ k} } $ is the unit vector on the Fermi surface, $v_F$ is the Fermi velocity,
$ \omega_n = (2n+1)\pi T $ is the Matsubara frequency, $n$ is an integer number, 
and $T$ is a temperature. In this paper, the symbol $ \hat{\cdots } $ 
represents $ 2 \times 2 $ matrix structure in spin space and $ \hat{\sigma}_j $ for 
$ j = 1$-$3 $ are the Pauli matrices. 
The vector potential is denoted by 
$ \boldsymbol{A}$ and the magnetic field $\boldsymbol{H}=\nabla \times \boldsymbol{A}$ 
is in the $z$ direction.
We introduced a definition 
$\undertilde{X}(\boldsymbol{r},{\boldsymbol{k}},i\omega_n)
\equiv X^\ast(\boldsymbol{r},-{\boldsymbol{k}},i\omega_n)$
for all functions $X$.
The electric current is given by
\begin{align}
 \boldsymbol{j}(\boldsymbol{r})=&
  \frac{\pi e v_F N_0}{2i} T \sum_{\omega_n} \int \frac{ d \boldsymbol{ {k} } }{2 \pi }
 \textrm{Tr} \left[ \check{T}_3\; \boldsymbol{k}\; \check{g}( \boldsymbol{r}, \boldsymbol{ {k} }, \omega_n )\right],
 \label{eq:j_qp}
\end{align}
with $\check{T}_3=\text{diag}[\hat{\sigma}_0,-\hat{\sigma}_0]$, where $\hat{\sigma}_0$ is the identity 
matrix in spin space and $N_0$ is the density of state per spin at the Fermi level.
We mainly consider the two equal-time pairing order parameters in two dimension:
spin-singlet $d$-wave symmetry $\hat{\Delta}(\boldsymbol{r},\theta)
=\Delta(\boldsymbol{r})\cos(2\theta) i\hat{\sigma}_2$ and 
spin-triplet $p$-wave symmetry $\hat{\Delta}(\boldsymbol{r},\theta)
=\Delta(\boldsymbol{r})\cos(\theta) \hat{\sigma}_1$, where $\theta$ is a directional angle
 with $k_x=\cos\theta$ and $k_y=\sin\theta$. 
The pair potentials are determined self-consistently from the gap equation 
\begin{gather}
 {\Delta} ( \boldsymbol{r} ) i\hat{\sigma}_\nu\hat{\sigma}_2=
  \pi N_0 g T \sum_{\omega_n } \int_0^{2\pi} \frac{ d \theta }{2 \pi }
 \hat{f}( \boldsymbol{r}, \theta, i\omega_n ) V_x(\theta),
 \label{eq:gap}
\end{gather}
where $x=s, p$ and $d$ indicate the pairing symmetry, $\nu=0$ and 3 for the spin-singlet and the spin-triplet order parameters, respectively.
The coupling constant $g$ satisfies
$\{N_0 g\}^{-1} =
   \ln ({T}/{T_c}) + \sum_{0 \le n < \omega_c / 2 \pi T } (n+1/2)^{-1}$
with $T_c$ and $\omega_c$ being the transition temperature and the cut-off energy, respectively.
The attractive potentials depends on the pairing symmetry $V_x(\theta)= s_x \phi_x(\theta)$ 
with $s_s=1$ and $\phi_s(\theta)=1$ for $s$ wave symmetry, $s_p=2$ and $\phi_p(\theta)=\cos\theta$ 
for $p$ wave symmetry, and $s_d=2$ and $\phi_d(\theta)=\cos(2\theta)$ for $d$ wave symmetry.
%
\begin{figure}[bt]
  \includegraphics[width=0.48\textwidth]{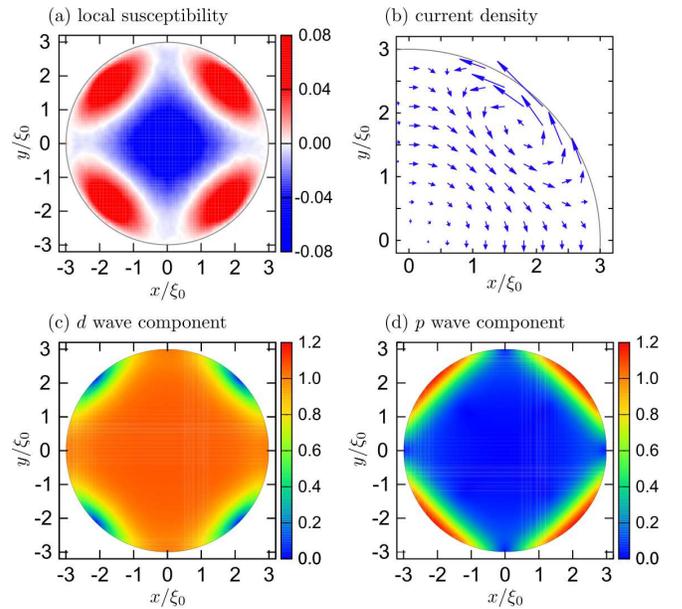}
  \caption{(a) The local susceptibility and (b) the current density of the $d$\,-wave 
           superconductor, where $R=3\,\xi_0$, $\lambda_L=3\,\xi_0$, $\omega_c=10\,\Delta_0$, 
           and $H^\mathrm{ext}= 0.001H_{c1}$. (c) The $d$-wave and (d) the $p$\,-wave components
           of the anomalous Green function. }
  \label{fig:result_d}
\end{figure}
The local magnetic susceptibility is defined by
\begin{align}
\chi_m(\boldsymbol{r}) = \left( H(\boldsymbol{r})
- H^\mathrm{ext} \right) / (4 \pi H^\mathrm{ext}),
\end{align}
where $H^\textrm{ext}$ is the uniform external magnetic field in the $z$ direction.
The susceptibility of the whole disk is calculated to be $\chi=\int d\boldsymbol{r} \chi_m(\boldsymbol{r}) /(\pi R^2)$.
In the absence of spin-dependent potential, the spin structure of $\hat{\Delta}$ 
and that of $\hat{f}$ are always the same with each other. 
We use the standard Riccati parametrization~\cite{Scho1,Scho2,eschrig} to 
solve the Eilenberger equation Eq.~(\ref{eilenberger_m2}).
To obtain numerical solutions of the Riccati type differential equation in closed disks, 
we apply a method discussed in Ref.~\onlinecite{Naga1}. 
An initial value at a certain place in the closed system is necessary to solve the Riccati equation.
The obtained solution usually depends on the initial condition.
However, when we solve the equation along the long enough classical trajectory, 
the effects of the initial condition is eliminated. 
In numerical simulation, we increase the length of the trajectory until 
solutions do not depend on the initial conditions.
The vector potential $ \boldsymbol{A} $ is obtained by solving the Maxwell equation
$ { \nabla } \times \boldsymbol{H} = ( 4 \pi/ c) \boldsymbol{j}$ with Eq.~(\ref{eq:j_qp}).
We calculate self-consistent solutions of the vector potential and pair potential by solving 
the Maxwell equation and the Eilenberger equation simultaneously.
The anomalous Green function $\hat{f}(\boldsymbol{r},\theta,i\omega_n)$ is originally defined 
by the two annihilation operators 
of two electrons consisting of a Cooper pair. Therefore $\hat{f}(\boldsymbol{r},\theta,i\omega_n)$ 
must be antisymmetric under the interchange of the two electrons, which stems from the 
Fermi-Dirac statistics of electrons.
Such fundamental relation is represented by
\begin{align}
\hat{f}(\boldsymbol{r},\theta,i\omega_n)=
-[\hat{f}(\boldsymbol{r},\theta+\pi,-i\omega_n)]^{\textrm{T}},\label{fermi_dirac}
\end{align}
 where T represents the transpose of matrices. 

\section{results}
\begin{figure}[tb]
    \includegraphics[width=0.48\textwidth]{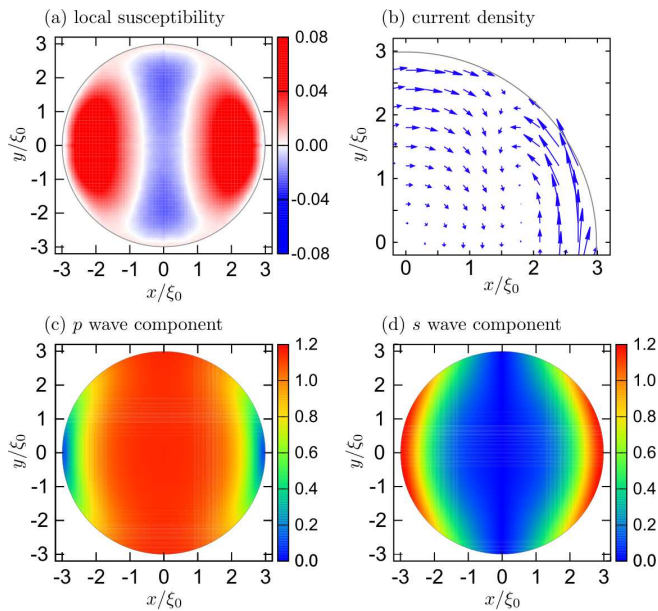}
  \caption{(a) The local susceptibility and (b) the current density of the $p$\,-wave 
               superconductor, where $R=3\,\xi_0$, $\lambda_L=3\,\xi_0$. 
           (c) The $p$\,-wave and (d) the $s$\,-wave component
               of the anomalous Green function.}
  \label{fig:result_p}
\end{figure}

The external magnetic field and the cut-off energy are fixed 
at $H^\mathrm{ext} = 0.001 H_{c_1}$ and $ \omega_c = 10 \Delta_0$, respectively.
Here $H_{c_1} = \hbar c/|e| \xi_0^2 $ is the first critical magnetic field.
The length is measured in units of $ \xi_0 = \hbar v_F / \Delta_0 $
with $\Delta_0$ being the amplitude of the pair potential at $ T=0 $.
The current density is normalized to $J_0 = \hbar c / |e|\xi_0^3 $.
The characteristic length scale of the Maxwell equation is $\lambda_L=
(4\pi n e^2/mc^2)^{-1/2}$ and is
a parameter in the numerical simulation.
Throughout this paper, we use a unit of $k_B=1$.

%
%

In Fig.~\ref{fig:result_d},
we first show the calculated results of the local susceptibility (a) and the current density (b)
for the $d$\,-wave superconducting disk, 
where we fix the parameters as $R = 3 \, \xi_0$, 
$\lambda_L=3 \, \xi_0$, and $ T= 0.3 \, T_c$. 
We set $+x$ and $+y$ axes to be identical to (100) and (010) directions of the high-$T_c$ crystal. 
The central region of the disk is diamagnetic as usual, whereas the surfaces in the $(110)$ 
and $(1\bar{1}0)$ directions are paramagnetic as shown in (a). 
The current density has the complex structure near the surface as shown in (b),
where the arrow indicates the direction of current and its length represents the amplitude
of current. Here we present the picture only for $x>0$ and $y>0$ in (b) because 
the results are fourfold symmetric due to the $d$\,-wave character of order parameter. 
The diamagnetic current flows at the edges in the (100) and (010)
directions, whereas the paramagnetic current flows at the edges in the 
(110) direction. The vortex-like current profile can be seen near the surfaces 
because the two currents flow the opposite directions to each other. 
At the central region, on the other hand, only the diamagnetic current flows.
Such magnetic properties are unique to unconventional superconductors.
In $s$\,-wave case, the susceptibility is diamagnetic everywhere in the 
disk as show in the Appendix~A.
\par

The anomalous paramagnetic response is well explained by appearing 
the odd-frequency Cooper pairs. 
The anomalous Green function can be decomposed into $s$\,-, $p$\,-, and $d$\,-wave components
by
\begin{align}
 {f}_x ( \boldsymbol{r},i\omega_n) i\hat{\sigma}_\nu\hat{\sigma}_2=&
 \int_0^{2\pi} \frac{ d \theta }{2 \pi } V_x(\theta)
 \hat{f}( \boldsymbol{r}, \theta, i\omega_n ),
\end{align}
for $x=s, p$, and $d$.
Figure \ref{fig:result_d}(c) shows the amplitude of the $d$\,-wave component at $\omega_0 = \pi T$. 
The spatial profile of the 
order parameter is almost similar to that of (c).
The $d$\,-wave 
component drastically suppresses in $(110)$ and $(1\bar{1}0)$ directions, which has been well known 
as a result of appearing of topologically protected Andreev surface bound states
at the zero-energy~\cite{hu,tanaka95}. At the same time, the $p$\,-wave component of the 
anomalous Green function grows at the corresponding edges as shown in (d).
The spin-singlet $p$\,-wave Cooper pairs must have the odd-frequency symmetry
to satisfy Eq.~(\ref{fermi_dirac}).
The breakdown of the translational symmetry at the surface mixes the even- and odd-parity components. 
The appearance of the Andreev surface bound states and that of the odd-frequency 
pairs are the two different faces of the same phenomenon. 
To have the zero-energy peak in the density of states, the frequency symmetry 
of Cooper pair must be odd~\cite{bergeret,tanaka07,Asa1}.
The odd-frequency pairs have so called \textsl{negative pair density}~\cite{Asa1},
which leads to the paramagnetic instability as shown in Appendix~B. Therefore we conclude that the 
paramagnetic current is carried by the induced odd-frequency Cooper pairs.
Comparing the Figs.~2(b) with 2(d), the paramagnetic current 
flows at the regions where the odd-frequency Cooper pairs stay.

We have also obtained qualitatively the same results for a spin-triplet $p$\,-wave 
superconducting disk at $T=0.2T_c$
as shown in Fig.~\ref{fig:result_p}, where 
 the local magnetic susceptibility (a), the current density (b), 
the $p$\,-wave component of $\hat{f}$ (c), and $s$\,-wave component of $\hat{f}$ (d) are 
presented in the same manner 
as Fig.~2. 
The results in Fig.~\ref{fig:result_p} show the twofold symmetry reflecting the $p$\,-wave
order parameter.
The surface bound states are appear at the $(100)$ surfaces at which
the $p$\,-wave component of the anomalous Green function is suppressed. 
Correspondingly, the $s$\,-wave component becomes large at the surfaces of (100) directions. 
The spin-triplet $s$\,-wave component belongs to the odd-frequency symmetry class
according to Eq.~(\ref{fermi_dirac}).
The main difference between Figs.~2 and 3 is the property of the surface ABS 
at the zero-energy. In the spin-triplet $p$\,-wave disk,  
Majorana fermions appear at the surface~\cite{asano13}.
From Figs.~2 and 3, we conclude that the magnetic property of
 unconventional superconductors are intrinsically inhomogeneous 
and can be paramagnetic 
because of the odd-frequency Cooper pairs at the surface. 

\par
\begin{figure}[!t]
  \includegraphics[width=0.4\textwidth]{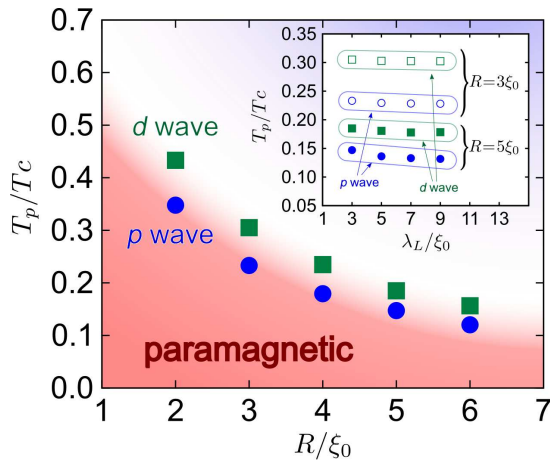}
  \caption{(a) The diamagnetic-paramagnetic phase diagram of superconducting disks 
               for the $d$\,-wave (square) and $p$\,-wave (circle) pairing symmetry, 
               where $\lambda_L=3\,\xi_0$, $\omega_c=10 \Delta_0$. 
           (b) The penetration length dependencies of the paramagnetic-diamagnetic 
               crossover temperatures, where the square and circle symbols are the results 
               for the $d$\,-wave and $p$\,-wave pairings, 
               and the open and closed symbols are the results 
               for the $R=3\,\xi_0$ and $R=5\,\xi_0$ superconducting disks, respectively.
               }
  \label{fig:phase diagram}
\end{figure}
%

The magnetic properties of superconductors 
strongly depends on the disk size because the odd-frequency pairs spatially localize near the 
surface limited by $\xi_0$ from the edge. Next, therefore, we discuss the relation between the 
magnetic property and the disk size.
Figure \ref{fig:phase diagram} is the paramagnetic-diamagnetic phase diagram of the $d$-\, 
and $p$\,-wave superconducting disks, 
where the vertical axis is the paramagnetic-diamagnetic crossover temperature $T_p$ 
and the horizontal one is the radius of superconducting disk $R$. 
The disk is paramagnetic $\chi >0$ at the temperatures below $T_p$.
The results show that
$T_p$ decrease with increasing the radius of the superconductor. 
As shown in Figs.~2 and 3, the paramagnetic area is limited to $\xi_0$ from the surface 
because odd-frequency pairs are confined there. On the other hand, the bulk area 
are diamagnetic because even-frequency pairs stay there.
Roughly speaking, the relative area of staying the odd-frequency pairs 
to the whole area of disk qualitatively determines the magnetic response of the disk.
Therefore the paramagnetic phase disappears in large disks with $R \gg \xi_0$.
because the contribution from the surface is negligible in large enough disks.
This argument is supported by the $\lambda_L$ dependence of $T_p$
shown in the inset of Fig.~4, 
where open (filled) symbols 
represent the results for $R/\xi_0=3$ (5) and the circles (squares) 
are the results for $p$ ($d$) wave disks. 
The crossover temperature is totally insensitive to $\lambda_L$. 
To be paramagnetic, the larger disks require the stronger contribution from the 
odd-frequency Cooper pairs.
The odd-frequency Cooper pairs energetically localize around the zero-energy~\cite{Asa1}. 
The temperature smears effects of them on the magnetic response.
Therefore $T_p$ decreases with increasing the disk size as shown in Fig.~4.

\par

Finally, we discuss the susceptibility of whole superconducting disk as a function of temperature
as shown in Fig.~5,
where we fix the penetration depth at $\lambda_L = 3 \, \xi_0$. 
The results for $d$\,- and $p$\,- wave symmetries are presented in (a) and (b), respectively.
The magnetic susceptibility just below $T_c$ is negative as usual.
With decreasing temperature, the paramagnetic current due to the odd-frequency 
Cooper pairs increases. As a consequence, the susceptibility upturns at low temperature,
which is qualitatively different from the susceptibility in the $s$ wave 
case as shown in Appendix~A. 
Below $T_p$, the paramagnetic odd-frequency Cooper pairs dominate the magnetic 
response of the superconductor. Therefore the dependence of the susceptibility 
on temperature shows the reentrant behavior as demonstrated in Fig.~5.
In experiments, it is possible to measure the susceptibility as a function of temperature.

\begin{figure}[!b]
    \includegraphics[width=0.48\textwidth]{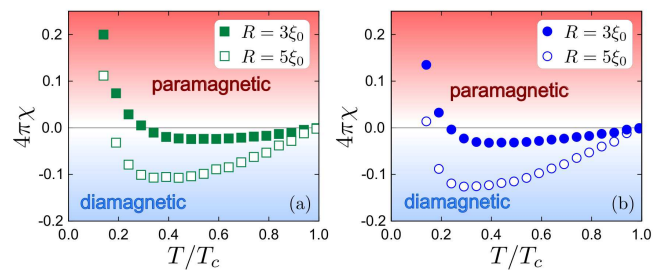}
  \caption{The temperature dependencies of the magnetic susceptibility 
           for the (a) $d$\,-wave and (b) $p$\,-wave superconducting disks.}
  \label{fig:tdep_chi}
\end{figure}

\section{Discussion}
Our theoretical results may correlate to
the measurement of the pair density at low temperature~\cite{walter}.
They measured the penetration depth $\lambda_L= (4 \pi n_se^2/mc^2)^{-1/2}$ 
of a YBCO film on which 
(110) oriented internal surfaces are introduced by heavy-ion bombardment.
They found that $\lambda$ first decreases with decreasing temperature from $T_c$ 
then increases at very low temperature.
This results can be interpreted as a result of decreasing the pair density $n_s$ 
at low temperature. The odd-frequency pairs have the negative pair density. Thus 
the decrease of $n_s$ may suggest the increase of odd-frequency pair fraction.
The experimental results on a high-$T_c$ superconductor are 
consistent with our theoretical results.

In real materials, the inelastic scatterings dephase the Cooper pairs 
and broaden the energy profile of the pairing functions. 
The inelastic mean free path also limits the size of disks in the phase diagram
shown in Fig.~4.
In $d$\,-wave superconductors, it has been shown that
the surface roughness also broadens the zero-energy peak at the surface. 
In such situation, we infer that the roughness would suppress the paramagnetic effect.
On the other hand in $p$\,-wave superconductors, the surface zero-energy peak is 
robust under the disordered potential. 
Therefore effects of surface roughness on the paramagnetic effect 
would be different in the two pairing symmetries. This is an important future issue.

The diamagnetism of superconductor is a result of gaining the condensation energy
below the transition temperature. Therefore the paramagnetic superconducting states 
may be impossible in uniform thermodynamic limit. The paramagnetic phase in Fig.~4 
can be considered as an unstable state and should disappear for large $R/\xi_0$. 
 As shown in Fig.~2. and 3, the magnetic inhomogeneity is an intrinsic feature of 
unconventional superconductors. Such inhomogeneous property assists the appearance
of the paramagnetic phase in small disks.
Indeed we confirm that the paramagnetic phase appears in two cooling 
processes: field cool and zero-field cool. 

The spontaneously time-reversal symmetry (TRS) breaking states 
has been discussed in high-$T_c$ grains~\cite{black}. The subdominant component
of order parameter near the surface breaks TRS. 
The results in Fig.~2 also indicates the TRS breaking superconducting state even 
when we simply assume the pure $d$ wave order parameter.
We are thinking that the symmetry crossover from $d$ wave to TRS breaking $d+is$ 
might be possible in small samples. To prove this, however, we need to compare the free-energy
among possible symmetry states. This issue goes beyond the scope of this paper.

Odd-frequency pairs appear also in superconductor/ferromagnet proximity 
structures~\cite{bergeret}. When odd-frequency pairs are dominant in the 
ferromagnet~\cite{ya07sfs,braude,eschrig08}, the paramagnetic instability
may lead to spontaneous current there~\cite{melnikov12}.

\section{conclusion}
In conclusion, we have theoretically studied the magnetic response of small 
unconventional superconducting disks by using the quasiclassical Green function method.
We conclude that small unconventional 
superconductors can be paramagnetic at low temperature 
due to the appearance of odd-frequency Cooper pairs at their surface. 
The magnetic properties of unconventional superconductors are intrinsically 
inhomogeneous as a result of their topologically nontrivial nature. 
Our results show up such universal property of unconventional superconductivity.

\begin{acknowledgments}
The authors are grateful to Y.~Tanaka and S.~Higashitani for useful discussion.
This work was supported by the "Topological Quantum Phenomena" (No. 22103002) Grant-in Aid for 
Scientific Research on Innovative Areas from the Ministry of Education, 
Culture, Sports, Science and Technology (MEXT) of Japan.
\end{acknowledgments}
%
%

\appendix
\section{Results for $s$\,-wave disk}
\begin{figure}[b]
  \includegraphics[width=0.48\textwidth]{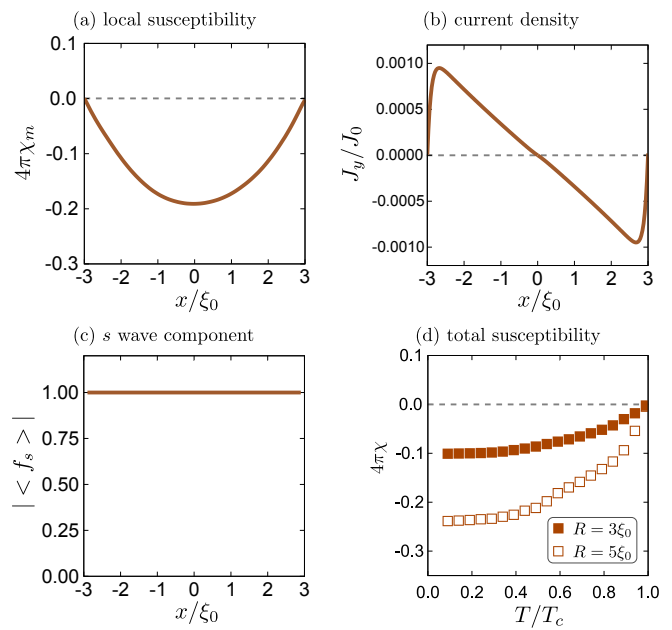}
  \caption{(a) The local susceptibility and (b) the current density of the $s$\, wave 
           superconductor, where $R=3\,\xi_0$, $\lambda_L=3\,\xi_0$, $\omega_c=10\,\Delta_0$, 
          $H^\mathrm{ext}= 0.001H_{c1}$ and $T=0.2T_c$. (c) The $s$\, wave component
           of the anomalous Green function. (d) The susceptibility vs temperature.}
\label{fig:swave}
\end{figure}
We supply the calculated results for conventional spin-singlet $s$\,-wave superconductors.
Fig.~\ref{fig:swave} shows the calculated results of the local susceptibility (a) and the current density (b)
for the $s$ wave superconducting disk, 
where we fix the parameters as $R = 3 \, \xi_0$, $\lambda_L=3 \, \xi_0$, and $ T= 0.2 \, T_c$. 
Because of the isotropic property in the $s$ wave pair potential, the results are also isotropic 
in real space. Therefore we plot the results as a function of $x$ at $y=0$.
The results in (a) show that the response is diamagnetic everywhere in the disk. Correspondingly 
the current profile in (b) suggests the usual Meissner screening current. The amplitude of the 
$s$\,-wave component of the anomalous Green function is almost uniform because $s$ wave superconductors
are topologically trivial and do not host any surface states. 
The amplitude of the $p$ wave component is much smaller than that of 
$s$ wave one.
The susceptibility of disk is plotted as a function of temperature in (d). 
The susceptibility decreases monotonically with decreasing temperature, which is usually observed 
in experiments.
Since the disk size is not much larger than $\lambda_L$, the perfect diamagnetism (i.e., $4 \pi \chi = -1$) is not archived. 

\section{Paramagnetic current due to odd-frequency pairs}
We discuss the contribution 
of odd-frequency pairs to the paramagnetic current 
 within the linear response theory.
Here we consider that
the pair potential has a single component in spin space $\hat{\Delta}(\boldsymbol{r},{\boldsymbol{k}})= {\Delta}(\boldsymbol{r},{\boldsymbol{k}})i\hat{\sigma}_\nu\hat{\sigma}_2$, where $\nu$ is one of $0$-$3$.
In such case, the Eilenberger equation is reduces to a $2\times 2$ matrix equation 
\begin{align}
&i\hbar v_F {\boldsymbol{k}} \cdot 
\boldsymbol{\nabla}_{\boldsymbol{r}} \, \hat{{g}} 
+\left[ \hat{H}, \hat{g} \right]=0,\\
&\hat{H}
=\left[\begin{array}{cc}
i\omega_n +\frac{ev_F}{c}{\boldsymbol{k}}\cdot \boldsymbol{A}& 
i{\Delta}(\boldsymbol{r},{\boldsymbol{k}}) \\
i\Delta(\boldsymbol{r},{\boldsymbol{k}}) & 
-i\omega_n -\frac{ev_F}{c}{\boldsymbol{k}}\cdot \boldsymbol{A}
\end{array}\right], \label{eilenberger}\\
&\hat{g}(\boldsymbol{r},{\boldsymbol{k}},\omega_n)
=\left[\begin{array}{cc}
{g}(\boldsymbol{r},{\boldsymbol{k}},\omega_n) & 
{f}(\boldsymbol{r},{\boldsymbol{k}},\omega_n) \\
 s_p\; \undertilde{f}(\boldsymbol{r},{\boldsymbol{k}},\omega_n) & 
-g(\boldsymbol{r},{\boldsymbol{k}},\omega_n)
\end{array}\right],\label{js}
\end{align}
where $s_p$ is 1 for even-parity order parameter and -1 for odd-parity one.
The electric current is given by
\begin{align}
\boldsymbol{j}(\boldsymbol{r})= - {2ie v_F\pi N_0}
T\sum_{\omega_n}\int\frac{d{\boldsymbol{k}}}{S_d}{\boldsymbol{k}}
  {g}(\boldsymbol{r},{\boldsymbol{k}},\omega_n).
\end{align}
Here $g$ is the normal Green function in the presence of the vector potential.
In what follows, we estimate $g$ within the linear response of $\boldsymbol{A}$.
In the Eilenberger equation, 
the vector potential formally shifts the energy.
Thus the Green function can be expressed as 
\begin{align}
g= g_0 + \partial_{\omega_n} g_0 (-iev_F/c) {\boldsymbol{k}}\cdot \boldsymbol{A}
\end{align}
within the linear response, where $g_0$ is the Green function at $\boldsymbol{A}=0$.
In what follows, we omitted "$0$" from the subscript of the Green function for simplicity.
By substituting the expression to Eq.~(\ref{js}), we obtain
\begin{align}
\boldsymbol{j}(\boldsymbol{r})=& - \frac{ n e^2 \pi \boldsymbol{A} }{2mc}
T\sum_{\omega_n}  
\partial_{\omega_n} \left\langle g(\boldsymbol{r},\omega_n) \right\rangle_{{\boldsymbol{k}}}\\
=& - \frac{n_s e^2 \boldsymbol{A}}{mc},\\
\frac{n_s}{n} =& \pi T\sum_{\omega_n} 
 \left\langle \partial_{\omega_n}
 {g}(\boldsymbol{r},\omega_n) \right\rangle_{{\boldsymbol{k}}}\\
=&\frac{1}{2} 
\int_{-\infty}^{\infty}d\omega \; 
 \left\langle \partial_{\omega}
 {g}(\boldsymbol{r},\omega) \right\rangle_{{\boldsymbol{k}}}
 \label{ns_m},
\end{align}
where $n$ is the density of electrons in the normal state and $N_0$ is the density of states 
per spin at the Fermi level. 
The $\omega_n$ derivative of the Green function can be defined only at $T=0$.
To discuss magnetic property of the anomalous Green function,
we write the derivative of $g$ to 
\begin{align}
 \partial_\omega g(\boldsymbol{r},{\boldsymbol{k}},\omega) =& 
 \frac{1}{2} \left( f_\text{E}^2 - f_\text{O}^2 \right) 
 \partial_\omega \ln \left( \frac{1+g}{1-g} \right),\label{ns_m2}\\
 f_\text{E}(\boldsymbol{r},{\boldsymbol{k}},\omega_n) = & 
 \left. \frac{1}{2}\left( f + s_p \undertilde{f} \right) \right|_{(\boldsymbol{r},{\boldsymbol{k}},\omega_n)},\\
  f_\text{O}(\boldsymbol{r},{\boldsymbol{k}},\omega_n) = & 
 \left. \frac{1}{2}\left( f - s_p \undertilde{f} \right) \right|_{(\boldsymbol{r},{\boldsymbol{k}},\omega_n)},
\end{align}
where we have used the normalization condition in the Matsubara 
representation $g^2+s_p f \undertilde{f}=1$~\cite{higashitani2}.

At $\boldsymbol{A}=0$, the Eilenberger equation is decomposed into three equations 
for the three components in the matrix structure,
\begin{align}
\hbar v_F {\boldsymbol{k}} \cdot \boldsymbol{\nabla}_{\boldsymbol{r}} \, g ~
=& 2\Delta f_{\text{O}}, \label{e_g}\\
\hbar v_F {\boldsymbol{k}} \cdot \boldsymbol{\nabla}_{\boldsymbol{r}} \, f_{\text{E}} 
=&  - 2\omega_n f_{\text{O}}, \label{e_fp}\\
\hbar v_F {\boldsymbol{k}} \cdot \boldsymbol{\nabla}_{\boldsymbol{r}} \, f_{\text{O}} 
=& 2 (\Delta g -  \omega_n f_{\text{E}} ).
\label{e_fm}
\end{align}
Here we note that the all functions are real when we delete the superconducting phase.
For the uniform case, we obtain the solution
\begin{align}
g= \frac{\omega_n}{\sqrt{\omega_n^2+\Delta^2}}, \quad f=s_p\undertilde{f}= 
f_{\text{E}}= \frac{\Delta}{\sqrt{\omega_n^2+\Delta^2}},
\end{align}
and $\quad f_{\text{O}}=0$.
In the uniform bulk region, $f_{\text{E}}$
is the source of the order parameter $\Delta$. Thus parity, spin, and frequency symmetries of
$f_{\text{E}}$ and those of $\Delta$ should be identical to each other.
The contribution of $f_{\text{E}}$ to the pair density must be positive in Eq.~(\ref{ns_m2})
because the uniform superconductor is diamagnetic.
The surface and the interface are source of the inhomogeneity 
in superconductor and mix the two orbital symmetry: even parity and odd parity. 
Nonzero spatial derivative in Eqs.~(\ref{e_g})$-$(\ref{e_fm}) allows $f_{\text{O}}$ 
component. 

The component $f_{\text{O}}$ is an odd function of $\omega_n$ 
as shown in Eq.~(\ref{e_g}) because
$g$ is always an odd function of $\omega_n$ due to a symmetry relationship 
$g(\boldsymbol{r}, {\boldsymbol{k}},\omega_n)=- g^\ast(\boldsymbol{r}, {\boldsymbol{k}},-\omega_n)$.
Equation (\ref{e_fp}) indicates that the frequency symmetry of $f_{\text{O}}$ are always opposite to 
those in $f_{\text{E}}$.
Therefore inhomogeneity induces $f_{\text{O}}$ component which has the odd-frequency symmetry.
Eq.~(\ref{ns_m2}) tells us that the pair density of odd-frequency component is negative, 
which leads to the paramagnetic response.


\begin{thebibliography}{99}
\bibitem{tinkham} See, for example, M.\ Tinkham, \textit{Introduction to Superconductivity}, 2nd ed., (McGraw-Hill, 1996). 

\bibitem{Thom1} D.\ J.\ Thompson, M.\ S.\ M.\ Minhaj, L.\ E.\ Wenger, and J.\ T.\ Chen, 
                Phys. Rev. Lett. \textbf{75}, 529 (1995).
\bibitem{Geim1} A.\ K.\ Geim, S.\ V.\ Dubonos, J.\ G.\ S.\ Lok, M.\ Henini and J.\ C.\ Maan, 
                Nature (London) \textbf{396}, 144 (1998).


\bibitem{Brau1} W.\ Braunisch, N.\ Knauf, V.\ Kataev, S. Neuhausen, A. Grutz, A. Kock, B. Roden,
                D.\ Khomskii, and D.\ Wohlleben, Phys. Rev. Lett. \textbf{68}, 1908 (1992).


\bibitem{Schl1} B. Schliepe, M. Stindtmann, I. Nikolic,  and K. Baberschke, 
                Phys. Rev. B \textbf{47}, 8331 (1993).				

\bibitem{walter} H.\ Walter, W.\ Prusseit, R.\ Semerad, H.\ Kinder, W.\ Assmann, H.\ Huber,
				H.\ Burkhardt, D.\ Rainer, and J.\ A.\ Sauls, Phys. Rev. Lett. \textbf{80}, 3598 (1998).

\bibitem{Visa1} P.\ Visani, A.\ C.\ Mota, and A.\ Pollini, Phys. Rev. Lett. \textbf{65}, 1514 (1990).

\bibitem{Mota1} A.\ C.\ Mota, P.\ Visani, A.\ Pollini, K.\ Aupke, Physica B \textbf{197}, 95 (1994). 

%
\bibitem{koshelev} A.\ E.\ Koshelev and A.\ I.\ Larkin, Phys. Rev. B \textbf{52}, 13559 (1995).

\bibitem{moshcalkov} V.\ V.\ Moshchalkov, X.\ G.\ Qiu, and V.\ Bruyndoncx, 
	Phys. Rev. B \textbf{55}, 11793 (1997).

\bibitem{dominguez} D.\ Dominguez, E.\ A.\ Jagla, and C.\ A.\ Balseiro,
	Phys. Rev. Lett. \textbf{72}, 2773 (1994).
	

\bibitem{buchholtz} L.~J.~Buchholtz and G.~Zwicknagl, Phys. Rev. B \textbf{23}, 5788 (1981). 

\bibitem{hara} J.\ Hara and K.\ Nagai, Prog. Theor. Phys. \textbf{74}, 1237 (1986).

\bibitem{hu} C.\ R.\ Hu, Phys. Rev. Lett. \textbf{72}, 1526 (1994).

\bibitem{tanaka95} Y.\ Tanaka and S.\ Kashiwaya, Phys.\ Rev.\ Lett.\ \textbf{74}, 3451 (1995).

\bibitem{sato} M.\ Sato, Y.\ Tanaka, K.\ Yada, T.\ Yokoyama, Phys. Rev. B \textbf{83}, 224511 (2011).
The topological number for two-dimensional $d$- and $p$\,-wave superconductors is defined 
in terms of the subgap wave functions at partial Brillouin zone in one-dimension. 
Such characterization explain the dispersionless zero-energy bound states at their surface 
in the clean limit.

\bibitem{yip} S.\ K.\ Yip and J.\ A.\ Sauls, Phys. Rev. Lett. \textbf{69}, 2264 (1992).

\bibitem{Zare1}  A.\ Zare, T.\ Dahm and N.\ Schopohl, Phys. Rev. Lett. \textbf{104}, 237001 (2010).


\bibitem{fogelstrom} M.\ Fogelstr\"{o}m, D.\ Rainer, and J.\ A.\ Sauls, Phys. Rev. Lett. \textbf{79}, 281 (1997).

\bibitem{barash} Yu.\ S.\ Barash, M.\ S.\ Kalenkov, and J.\ Kurkijarvi, Phys. Rev. B \textbf{62}, 6665 (2000).

%
\bibitem{Hig1}   S.\ Higashitani, J. Phys. Soc. Jpn. \textbf{66}, 2556 (1997).

\bibitem{lofwander} T.\ L\"{o}fwander, V.\ S.\ Shumeiko, and G.\ Wendin,
Phys. Rev. B \textbf{62}, 14653R (2000).

\bibitem{AGD}     A.\ A.\ Abrikosov, L.\ P.\ Gor'kov, I.\ E.\ Dzyaloshinski,
\textit{Methods of Quantum Field Theory in Statistical Physics} (Dover Publications ,1975).  

\bibitem{belzig} W.\ Belzig, C.\ Bruder, and A. L. Fauchere, Phys. Rev. B, \textbf{58}, 14531 (1998).


\bibitem{ya04} Y.\ Asano, Y.\ Tanaka, and S.\ Kashiwaya, Phys. Rev. B \textbf{69}, 134501 (2004).

\bibitem{Lutchyn2010}
R.\ M.\ Lutchyn, J.\ D.\ Sau, and S.\ Das Sarma, Phys. Rev. Lett. \textbf{105}, 077001 (2010). 


\bibitem{Oreg}
Y.\ Oreg, G.\ Refael, and F.\ von Oppen, Phys. Rev. Lett. \textbf{105}, 177002 (2010).


\bibitem{asano13} Y.\ Asano and Y.\ Tanaka, Phys. Rev. B \textbf{87}, 104513 (2013).

\bibitem{Asa1}   Y.\ Asano, A.\ A.\ Golubov, Y.\ V.\ Fominov, and Y.\ Tanaka
                 Phys. Rev. Lett. \textbf{107}, 087001 (2011).

%
\bibitem{yoko1}  T.\ Yokoyama, Y\ Tanaka, N\ Nagaosa, 
                 Phys. Rev.Lett. \textbf{106}, 246601 (2011).

\bibitem{melnikov12} S.\ Mironov, A.\ Mel'nikov, and A.\ Buzdin, 
                     Phys. Rev. Lett. \textbf{109}, 237002 (2012).






\bibitem{Eilen}  G.\ Eilenberger, Z. Phys. \textbf{214}, 195 (1968).

\bibitem{Scho1}  N.\ Schopohl and K.\  Maki, Phys. Rev. B \textbf{52}, 490 (1995).
\bibitem{Scho2}  N.\ Schopohl, arXiv:cond-mat/9804064.

\bibitem{eschrig} M.\ Eschrig, Phys. Rev. B \textbf{80}, 134511 (2009).
%
%
\bibitem{Naga1}  Y.\  Nagai, K.\ Tanaka, and N.\ Hayashi, 
                 Phys. Rev. B \textbf{86}, 094526 (2012).

\bibitem{tanaka07} Y.\ Tanaka and A.\ A.\ Golubov, Phys. Rev. Lett. \textbf{98}, 037003 (2007).

\bibitem{bergeret} F.~S.~Bergeret, A.~F.~Volkov, and K.~B.~Efetov,
Phys. Rev. Lett. \textbf{86}, 4096 (2001).


\bibitem{ya07sfs} Y.\ Asano, Y.\ Tanaka, and A.~A.\ Golubov, Phys.\ Rev.\ Lett. \textbf{98}, 107002 (2007).

\bibitem{braude} V.\ Braude and Yu.~V.\ Nazarov, Phys.\ Rev.\ Lett.\ \textbf{98}, 077003 (2007).

\bibitem{eschrig08} M.\ Eschrig and T.\ L\"{o}fwander, Nat. Phys.\ \textbf{4}, 138 (2008).

\bibitem{black} A.\  M. Black-Schaffer, D.\ S.\ Golubev, T.\ Bauch, F.\  Lombardi, and M.\  Fogelstr\"{o}m,
Phys.\ Rev.\ Lett.\ \textbf{110}, 197001 (2013).


\bibitem{higashitani2} S.\ Higashitani, Phys. Rev. B \textbf{89}, 184505 (2014).

\end{thebibliography}
\end{document}